\def\year{2015}
\documentclass{sig-alternate}
\usepackage{times}
\usepackage{helvet}
\usepackage{courier}

\usepackage[colorlinks=true,linkcolor=blue]{hyperref}	
\usepackage{caption}
\usepackage{subcaption}
\usepackage{bbm}

\usepackage{CJKutf8} 
\usepackage{ucs} 
\usepackage[encapsulated]{CJK} 
\newcommand{\myfont}{gbsn}
\usepackage[usenames]{color}

\usepackage{amssymb,amsmath}
\usepackage{graphicx}
\begin{document}

\title{Learning to Discover Key Moments in Social Media Streams}
%
%
%
%
%

\numberofauthors{3} 
%
\author{
%
%
\alignauthor
Cody Buntain\\
       \affaddr{Dept. of Computer Science}\\
       \affaddr{University of Maryland}\\
       \affaddr{College Park, Maryland 20742}\\
       \email{cbuntain@cs.umd.edu}
\alignauthor
Jimmy Lin\\
       \affaddr{College of Information Studies}\\
       \affaddr{University of Maryland}\\
       \affaddr{College Park, Maryland 20742}\\
       \email{jimmylin@cs.umd.edu}
\alignauthor
Jennifer Golbeck\\
       \affaddr{College of Information Studies}\\
       \affaddr{University of Maryland}\\
       \affaddr{College Park, Maryland 20742}\\
       \email{golbeck@cs.umd.edu}
}

\maketitle
\begin{abstract}
This paper introduces LABurst, a general technique for identifying key moments, or moments of high impact, in social media streams without the need for domain-specific information or seed keywords.
We leverage machine learning to model temporal patterns around bursts in Twitter's unfiltered public sample stream and build a classifier to identify tokens experiencing these bursts.
We show LABurst performs competitively with existing burst detection techniques while simultaneously providing insight into and detection of unanticipated moments.
To demonstrate our approach's potential, we compare two baseline event-detection algorithms with our language-agnostic algorithm to detect key moments across three major sporting competitions: 2013 World Series, 2014 Super Bowl, and 2014 World Cup.
Our results show LABurst outperforms a time series analysis baseline and is competitive with a domain-specific baseline even though we operate without any domain knowledge.
We then go further by transferring LABurst's models learned in the sports domain to the task of identifying earthquakes in Japan and show our method detects large spikes in earthquake-related tokens within two minutes of the actual event.
\end{abstract}

%

\section{Introduction}

Though researchers have presented many methods for $\:$ \mbox{adapting} social media streams into news sources for journalists or first responders, many current approaches rely on prior knowledge and manual keyword engineering to detect events of interest.
While straightforward and capable, such approaches are often constrained to events one can easily anticipate or describe in very general terms, potentially missing impactful but \emph{unexpected} key moments.
For instance, one can follow the frequency of words like ``goal'' on Twitter during the 2014 World Cup to detect when goals are scored \cite{Cipriani2014}, but interesting occurrences like penalties or missed goals would be missed.
One might respond to this weakness by tracking additional penalty-related tokens, but this approach is untenable in that one cannot continually enlarge the keyword set.
Furthermore, one would still be unable to identify an unexpected moment like Luis Suarez's biting Giorgio Chiellini during the Uruguay-Italy World Cup match; who would have thought to include ``bite'' as a relevant token during that event?
Relying on predefined keywords also restricts these systems to those languages represented in the seed keyword set, a significant issue for international events like the World Cup.

Given the sheer volume of social media data (hundreds of thousands of comments, statuses, and photos are generated per minute on Facebook alone as of 2012 \cite{Pring2012}), one could instead forgo seed keywords completely and leverage time series analysis to track bursts in message volume (as with Vasudevan et al. \cite{vasudevan2013twitter}).
Such methods gain flexibility of domain but sacrifice semantic information about detected events (as one would need to extract keywords causing such bursts manually).
In this paper, we propose leveraging machine learning to combine both techniques.

To explore this integration, we introduce LABurst (for language-agnostic burst detection), a general  method to model bursts in token usage in social media streams.
The volume of these bursts then indicate the presence of a high-impact occurrence or key moment. 
In short, the more tokens experiencing a simultaneous burst, the higher the impact of that moment.
Contrasting with existing work, our approach is a streaming algorithm for \textbf{unfiltered} social media streams that discovers high-impact moments without prior knowledge of the target event and yields a description of the discovered moment.
Illustrating this flexibility is a collection of experiments on Twitter's sample stream surrounding key moments in large sporting competitions and natural disasters.
These experiments compare LABurst to two existing burst detection methods: a time series-based burst detection technique, and a domain-specific technique with a pre-determined set of sports-related keywords.
Results from these experiments demonstrate LABurst's competitiveness with existing methods. 

This work makes the following contributions:
\begin{itemize}
\item Presents a \textbf{streaming} algorithm and feature set for the discovery and description of impactful and unexpected key moments in Twitter's public sample stream \textbf{without requiring manually-defined keywords} as input,

\item Demonstrates our approach's performance is both \textbf{competitive}  and \textbf{flexible}, and

\item \textbf{Transfers} sports-trained models to \textbf{disaster response} with comparable performance.
\end{itemize}

\section{Related Work}
\label{sect:relatedWork}

Though LABurst focuses on the slightly different problem of discovering interesting moments in social media streams, our work shares foundations with classical event detection research.
Identifying key events from the ever-growing body of digital media has fascinated researchers for over twenty years, starting from digital newsprint to blogs and now social media \cite{allan1998line}.
Early event detection research followed that of Fung et al. in 2005, who built on the burst detection scheme presented by Kleinberg by identifying bursty keywords from digital newspapers and clustering these keywords into groups to identify bursty events \cite{Kleinberg:2002:BHS:775047.775061,Fung:2005:PFB:1083592.1083616}.
This work succeeded in identifying trending events and showed such detection tasks are feasible.
Recognizing that newsprint differs substantially from social media both in content and velocity, the research community began experimenting with new social media sources like blogs, but real gains came when microblogging platforms began their rise in popularity.
These microblogging platforms include Twitter and Sina Weibo and are characterized by constrained post sizes (e.g., Twitter constrains user posts to 140 characters) and broadcasting publicly consumable information.

One of the most well-known works in detecting events from microblog streams is Sakaki, Okazaki, and Matsuo's 2010 paper on detecting earthquakes in Japan using Twitter \cite{Sakaki:2010:EST:1772690.1772777}.
Sakaki et al. show that not only can one detect earthquakes on Twitter but also that it can be done simply by tracking frequencies of earthquake-related tokens.
Surprisingly, this approach can outperform geological earthquake detection tools since digital data propagates faster than tremor waves in the Earth's crust.
Though this research is limited in that it requires pre-specified tokens and is highly domain- and location-specific (Japan has a high density of Twitter users, so earthquake detection may perform less well in areas with fewer Twitter users), it demonstrates a significant use case and the potential of such applications.

Along with Sakaki et al., 2010 saw two other relevant papers: Lin et al.'s construction of a probabilistic popular event tracker \cite{Lin:2010:PSM:1835804.1835922} and Petrovi\'{c}, Osborne, and Lavrenko's application of locality-sensitive hashing (LSH) for detecting first-story tweets from Twitter streams \cite{Petrovic:2010:SFS:1857999.1858020}.
Lin's work demonstrated that the integration of non-textual social and structural features into event detection could produce real performance gains.
Like many contemporary systems, however Lin's models require seeding with pre-specified tokens to guide its event detection and concentrates on retrospective per-day topics and events.
In contrast, Petrovi\'{c} et al.'s clustering research in Twitter avoids the need for seed keywords and retrospective analysis by instead focusing on the practical considerations of clustering large streams of data quickly.
While typical clustering algorithms require distance calculations for all pairwise messages, LSH facilitates rapid clustering at the scale necessary to support event detection in Twitter streams by restricting the number of tweets compared to only those within some threshold of similarity.
Once these clusters are generated, Petrovi\'{c} was able to track their growth over time to determine impact for a given event.
This research was unique in that it was one of the early methods that did not require seed tokens for detecting events and has been very influential, resulting in a number of additional publications to demonstrate its utility in breaking news and for high-impact crisis events \cite{osborne2014real,petrovic2013can,6601695}.
Petrovi\'{c}'s work and related semantic clustering approaches rely on textual similarity between tweets, which limits its ability to operate in mixed-language environments and differentiates LABurst and its language agnosticism.

Similar to Petrovi\'{c}, Weng and Lee's 2011 paper on EDCoW, short for Event Detection with Clustering of Wavelet-based Signals, is also able to identify events from Twitter without seed keywords \cite{weng2011event}.
After stringent filtering (removing stop words, common words, and non-English tokens), EDCoW uses wavelet analysis to isolate and identify bursts in token usage as a sliding window advances along the social media stream.
Besides the heavy filtering of the input data, this approach exhibits notable similarities with the language-agnostic method we describe herein with its reliance on bursts to detect event-related tokens.
These methods, however, operate retrospectively, focusing on daily news rather than breaking event detection on which our research focuses.
Becker, Naaman, and Gravano's 2011 paper on identifying events in Twitter also fall under retrospective analysis, but their findings also demonstrate reasonable performance in identifying events in Twitter by leveraging classification tasks to separate tweets into those on ``real-world events'' versus non-event messages \cite{becker2011beyond}.
Similarly, Diao et al. also employ a retrospective technique to separate tweets into global, event-related topics and personal topics \cite{diao2012finding}.

Many researchers have explored motivations for using platforms like Twitter and have shown interesting dynamics in our behavior around events with broad impact.
For instance, Lehmann et al.'s 2012 work on collective attention on Twitter explores hashtags and the different classes of activity around their use \cite{Lehmann:2012:DCC:2187836.2187871}.
Their work includes a class for activity surrounding unexpected, exogenous events, characterized by a peak in hashtag usage with little activity leading up to the event, which lends credence to our use of burst detection for identifying such events.
Additionally, this interest in burst detection has led to several domain-specific research efforts that also target sporting events specifically\cite{vasudevan2013twitter,Zhao2011,lanagan2011using}.
Lanagan and Smeaton's work is of particular interest because it relies almost solely on detecting bursts in Twitter's per-second message volume, which we use as inspiration for one of our baseline methods discussed below.
Though naive, this frequency approach is able to detect large bursts on Twitter in high-impact events  without complex linguist analysis and performs well in streaming contexts as little information must be kept in memory.
Detecting such bursts provide evidence of an event, but it is difficult to gain insight into that event  without additional processing.
LABurst addresses this need by identifying both the overall burst and keywords related to that burst.

More recently, Xie et al.'s 2013 paper on TopicSketch seeks to perform real-time event detection from Twitter streams ``without pre-defined topical keywords'' by maintaining acceleration features across three levels of granularity: individual token, bigram, and total stream \cite{xie2013topicsketch}.
As with Petrovi\'{c}'s use of LSH, Xie et al. leverage ``sketches'' and dimensionality reduction to facilitate event detection and also relies on language-specific similarities.
Furthermore, Xie et al. focus only on tweets from Singapore rather than the worldwide stream.
In contrast, our approach is differentiated primarily in its language-agnosticism and its use of the unfiltered stream from Twitter's global network.

Despite this extensive body of research, it is worth asking how event detection on Twitter streams differs from Twitter's own offerings on ``Trending Topics,'' which they make available to all their users.
When a user visit's Twitter's website, she is immediately greeted with her personal feed as well as a listing of trending topics for her city, country, worldwide, or nearly any location she chooses.
These topics offer insight into the current popular topics on Twitter, but the main differentiating factor is that these popular topics are not necessarily connected to specific events.
Rather, popular memetic content like ``\#MyLovelifeInMoveTitles'' often appear on the list of trending topics.
Additionally, Twitter monetizes these trending topics as a form of advertising \cite{Sydell2011}.  
These trending topics also can be more high-level than the interesting moments we seek to identify: for instance, during the World Cup, particular matches or the tournament in general were identified as trending topics by Twitter, but individual events like goals or penalty cards in those matches were not.
It should be clear then that Twitter's trending topics serves a different purpose than the streaming event detection described herein.

\section{Moment Discovery Defined}
\label{sect:model}

This paper demonstrates the LABurst algorithm's ability to discover and describe impactful moments from social media streams \emph{without} prior knowledge of the types or domains of these target moments.
To that end, we first lay LABurst's foundations by defining the problem LABurst seeks to solve and presenting the model around which LABurst is built.

\subsection{Problem Definition}

Given an unfiltered (though potentially down-sampled) $\;$ stream $S$ of messages $m$ consisting of various tokens $w$ $\;$ (where a ``token'' is defined as a space-delimited string)\footnote{Our use of ``token'' is more general than a ``keyword'' as it includes numbers, emoticons, hashtags, or web links}, our objective is to determine whether each time slice $t$ contains a impactful moment and, if so, extract tokens that describe the moment.
Identifying and describing such moments separately is difficult because, by the time one can react to a key moment with a separate analysis tool, the moment may have passed.
We define a ``key moment'' here as a brief instant in time, lasting on the order of seconds, that a journalist would label as ``breaking news.''
Key moments might comprise the highlights of a sporting competition or be the moment an earthquake strikes, the moment a terrorist attack occurrs, or similar.
Such moments often generate significant popular interest, affect large populations, or represent an otherwise instrumental moment in larger event (e.g., the World Cup).
By focusing on these instantaneous moments of activity, we also avoid the complexities of defining an ``event'' and the hierarchies among them.

Formally, we let $E$ denote the set of all time slices $t$ in which a key moment occurs.
The indicator function $\mathbbm{1}_E(S_t, t)$ takes the stream $S$ up to time $t$ and returns a $1$ for all times in which an impactful moment occurs, and $0$ for all other values of $t$. 
We then define the moment discovery task as approximating this indicator function $\mathbbm{1}_E(S_t, t)$.
We also include a function $B_E(S_t, t)$ that returns a set of words $w$ that describe the discovered moment at time $t$ if $t \in E$ and an empty set otherwise.
To account for possible lag in reporting the event, typing out a message about the event, and the message actually posting to a social media server, we include a delay parameter $\tau$.
This parameter relaxes the task by constructing the set $E'$ where, for all $t \in E$, $t, t+1, t+2, ..., t+\tau \in E'$.
Since our evaluation compares methods that share the same ground truth, and controlling $\tau$ affects the ground truth consistently, comparative results should be unaffected.
In this paper, we use $\tau=2$.

False positives/negatives and true positives/negatives follow in the normal way for some candidate function $\widehat{\mathbbm{1}_{E'}}(S_t, t)$: a false positive is any time $t$ such that $\widehat{\mathbbm{1}_{E'}}(S_t, t) = 1$ and ${\mathbbm{1}_{E'}}(S_t, t) = 0$; likewise, a false negative is any $t$ such that $\widehat{\mathbbm{1}_{E'}}(S_t, t) = 0$ and ${\mathbbm{1}_{E'}}(S_t, t) = 1$.
True positives/negatives follow as expected.

\subsection{The LABurst Model}

In LABurst, we sought to combine the language-agnostic flexibility of burst detection techniques with the specificity of domain-specific keyword burst detectors.
This integration results from ingesting a social media stream, maintaining a sliding window of frequencies for \textbf{each} token contained within the stream, and using the number of bursty tokens in a given minute as an indicator of the moment's impact.
Critically, these tokens can be of any language and are neither stemmed, normalized, or otherwise modified.
As an example, after a goal is scored in a World Cup match, one would expect to see many different forms of the word ``goal'' (both different languages and different variations, such as ``gooooaaal'') experiencing a burst within a minute of the score.
Most other approaches use language models to collapse these various token forms, whereas LABurst leverages this information as a predictor.

\begin{figure}[hbtp]
\begin{center}
\includegraphics[width=2.5in]{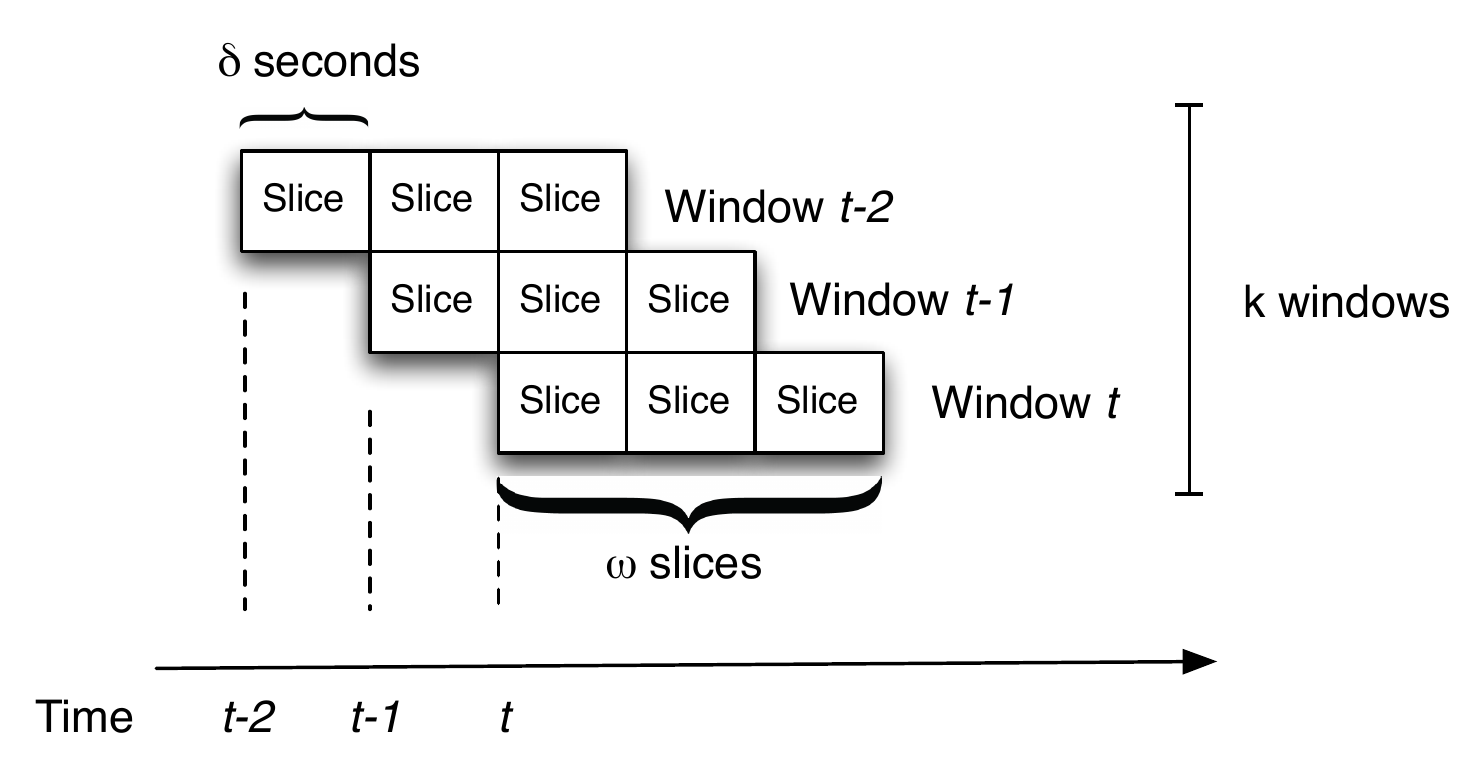}
\caption{LABurst Sliding Window Model}
\label{fig:windowSlices}
\end{center}
\end{figure}

At a lower level, LABurst runs a sliding window over the incoming data stream $S$ and divides it into slices of a fixed number of seconds $\delta$ such that time $t_{i} - t_{i-1} = \delta$.
LABurst then combines a set number $\omega$ of these slices into a single window (with an overlap of $\omega - 1$ slices), splits each message in that window into a set of tokens, and tabulates each token's frequency.
By maintaining a list of frequency tables from the past $k$ windows up to time $t$ (see Figure \ref{fig:windowSlices}), we construct features describing a token's changes in frequency.
From these features, we use machine learning to separate tokens into two classes: bursty tokens $\mathbf{B}_t$, and non-bursty tokens $\mathbf{B}'_t$.
Following this classification, if the number of bursty tokens exceeds some threshold $|\mathbf{B}_t| \ge \rho$, LABurst flags this window at time $t$ as containing a high-impact moment.
In this manner, LABurst approximates the target indicator function with $\widehat{\mathbbm{1}_{E'}}(S_t, t) = |\mathbf{B}_t| \ge \rho$ and yields $\mathbf{B}_t$ as the set of descriptive tokens for the given moment.

To avoid spurious bursts generated by endogenous network phenomena, retweets are discarded since existing literature shows retweets propagate extremely rapidly, leading to possible false bursts \cite{kwak2010twitter}.

\subsubsection{Temporal Features}

To capture token burst dynamics, we constructed a set of temporal and graphical features to model these effects, shown in Table \ref{tab:features}.
These features were calculated per token and normalized into the range $[0, 1]$ to avoid scaling issues.
Each feature's relative importance was then examined through an ablation study described later.\\

\begin{table}[h]
\caption{Features}
\begin{center}
\begin{tabular}{|p{1in}|p{2.0in}|}
\hline
\textbf{Feature} & \multicolumn{1}{|c|}{\textbf{Description}} \\ \hline
Frequency Regression	 & 	Given the logarithm of a token's frequency at each window, take the slope of the line that best fits this data. This feature is also duplicated for message frequency and user frequency as well.\\ \hline
Average Frequency Difference	 & 	The difference between the token's frequency in the most recent window and the average frequency across the previous $k -1$ windows. As with the regression feature, this feature was also calculated for message frequency and user frequency.\\ \hline
Inter-Arrival Time	 & 	The average number of seconds between token occurrences in the previous $k$ windows.\\ \hline
Entropy	 & 	The entropy of the set of messages containing a given token.\\ \hline
Density	 & 	The density of the @-mention network of users who use a given token.\\ \hline
TF-IDF	 & 	The term frequency, inverse document frequency for a each token.\\ \hline
TF-PDF	 & 	A modified version of TF-IDF called term frequency, proportional document frequency \cite{Bun:2002:TEN:645962.674082}.\\ \hline
BursT	 & 	Weight using a combination of a given token's actual frequency and expected token frequency \cite{Lee:2011:BDT:2009463.2009531}.\\ \hline
\end{tabular}
\end{center}
\label{tab:features}
\end{table}%

\subsubsection{LABurst's Bursty Token Classification}

LABurst's primary capability is its ability to differentiate between bursty and non-bursty tokens.
To make this determination, LABurst integrates these temporal features into feature vectors for each token and processes them using an ensemble of known classification algorithms.
Specifically, we use ensembles of support vector machines (SVMs) \cite{vapniksvm} and random forests (RFs) \cite{breiman2001random} integrated using AdaBoost \cite{Freund1995}.

Training these burst detection classifiers, however, requires both positive and negative samples of bursty tokens.
While obtaining positive samples of bursty tokens is relatively straightforward, negative samples are problematic.
For positive samples, we can identify high-impact, real-world events and construct a set of seed tokens that \emph{should} experience bursts along with the event (as done in typical seed-based event detection approaches).
Negative samples, however, are difficult to identify since one cannot know all events occurring around the world at a given moment.
To address this difficulty, we rely on a trick of linguistics and use stop words as negative samples, our justification being that stop words are in general highly used but used consistently (i.e., stop words are intrinsically non-bursty).
Therefore, in our experiments, we train LABurst on a set of events with known bursty tokens and stop words in both English and Spanish.
As this task is semi-supervised, we also include a self-training phase to expand our list of bursty tokens.

\section{Evaluation Framework}
\label{sect:methods}

Having established the details of our model, we now turn to frameworks for evaluating LABurst compared to existing methods.
To explore such comparisons, we first look to similar methods for detecting interesting events from social media streams and compare their performance relative to LABurst.
We then include a second experiment to demonstrate LABurst's domain independence and utility in the disaster response context.

\subsection{Accuracy in Event Discovery}

Our first research question is \textbf{RQ1}: is LABurst able to identify key moments as well as existing systems?
To answer this question, we constructed an experiment for enumerating key moments during major sporting competitions.
Such competitions are interesting given their large followings (many fans to post on social media), thorough coverage by sports journalists (high-quality ground truth), and regular occurrence (large volume of data), making them ideal for both data collection and evaluation.
Such events are also complex in that they include multiple types of events and unpredictable patterns of events around scores, fouls, and other compelling moments of play.

Our first step here was to collect data from a number of popular sporting events and identify key moments in each competition.
We captured these moments and their times from sports journalism articles, game highlights, box scores, blog posts, and social media messages.
These moments then comprise our ground truth.

We then introduced a pair of baseline methods: first, a time-series algorithm using raw message frequency following the approaches of Vasudevan et al. and the ``activity peak detection'' method set forth by Lehmann et al. \cite{vasudevan2013twitter,Lehmann:2012:DCC:2187836.2187871}, and second, a seed keyword-based algorithm in the pattern of Cipriani and Zhao et al. \cite{Cipriani2014,Zhao2011}.
We then evaluate the relative performance for LABurst and both baselines as described below.

\subsubsection{Sporting Competitions}

To minimize bias, these competitions covered several different sporting types, from horse racing to the National Football League (NFL), to F\'{e}d\'{e}ration Internationale de Football Association (FIFA) premier league soccer, to the National Hockey League (NHL), National Basketball Assoc. (NBA), and Major League Baseball (MLB). 
Each competition also contained four basic types of events: beginning of the competition, its end, scores, and penalties.
Table \ref{tab:eventStats} lists the events we identified and the number of key moments in each.
%
\begin{table}[htdp]
\footnotesize
\caption{Sporting Competition Data}
\begin{center}
\begin{tabular}{|p{2in}|c|c|c|}
\hline
\multicolumn{1}{|c|}{\textbf{Sport}} & \textbf{Key Moments} \\ \hline
\multicolumn{2}{|c|}{\textbf{Training Data}}  \\ \hline
2010 NFL Division Championship & 13 \\ \hline
2012 Premier League Soccer Games & 21 \\ \hline
2014 NHL Stanley Cup Playoffs & 24 \\ \hline
2014 NBA Playoffs & 3 \\ \hline
2014 Kentucky Derby Horse Race & 3 \\ \hline
2014 Belmont Stakes Horse Race & 3 \\ \hline
2014 FIFA World Cup Stages A+B & 80 \\ \hline

\multicolumn{2}{|c|}{\textbf{Testing Data}}  \\ \hline
2013 MLB World Series Game 5  & 7 \\ \hline
2013 MLB World Series Game 6  & 8 \\ \hline
2014 NFL Super Bowl & 13 \\ \hline
2014 FIFA World Cup Third Place & 11 \\ \hline
2014 FIFA World Cup Final & 7 \\ \hline
\multicolumn{1}{|r|}{\textbf{Total}} & 193 \\ \hline
\end{tabular}
\end{center}
\label{tab:eventStats}
\end{table}

In 2012, we tracked four Premier League games in November.
For the 2013 World Series between the Boston Red Sox and the St. Louis Cardinals, we covered the final two games on 28 October and 30 October of 2013.
Likewise, we tracked a subset of playoff games during the 2014 NHL Stanley Cup and NBA playoffs.
For the 2014 World Cup, our analysis included a number of early matches during stages 1 and 2 and the the final two matches of tournament: the 12 July match between the Netherlands and Brazil for third place, and the final match on 13 July between Germany and Argentina for first place.

These events were split into training and testing sets; training data covered the 2010 NFL championship, 2012 premier league soccer games, NHL/NBA playoffs, the Kentucky Derby/Belmont Stakes horse races, and several days of World Cup matches in June of 2014.
The testing data covered the 2013 MLB World Series, 2014 NFL Super Bowl, and the final two matches of the 2014 FIFA World Cup.

\subsubsection{Burst Detection Baselines}

The LABurst algorithm straddles the line between time-series analysis and token-centric burst detectors.
Therefore, to evaluate LABurst properly, we implemented two baselines for comparison.
The first baseline, to which we refer as RawBurst, uses a known method for detecting bursts by taking the difference between the number of messages seen in the current time slice and the average number of messages seen over the past $k$ time slices \cite{vasudevan2013twitter,Lehmann:2012:DCC:2187836.2187871}.

Formally, we define a series of time slices $t \in T$ segmented into $\delta$ seconds and a social media stream $S$ containing messages $m$ such that $S_t$ contains all messages in the stream between $t-1$ and $t$.
We then define the frequency of a given time slice $t$ as $\text{freq}(t, S) = |S_t|$ and the average over the past $k$ time slices as $\text{avg}(k, t, S)$, shown in Eq. \ref{eq:windowAverage}.
\begin{equation}
\label{eq:windowAverage}
\text{avg}(k, t, S) = \frac{\sum_{j=t-k}^{t}\text{freq}(j, S)}{k}
\end{equation}
Given these functions, we take the difference $\Delta_{t, k}$ between the frequency at time $t$ and the average over the past $k$ slices such that $\Delta_{t, k} = \text{freq}(t, S) - \text{avg}(k, t, S)$.
If this difference exceeds some threshold $\rho$ such that $\Delta_{t, k} \ge \rho$, we say an event was detected at time $t$.

Following those like Cipriani from Twitter's Developer Blog and others, we then modify the RawBurst algorithm to detect events using frequencies of a small set of seed tokens $w \in W$, to which we will refer as TokenBurst  \cite{Cipriani2014}. 
To convert RawBurst into TokenBurst, we modify the $\text{freq}(t, S)$ function to return the summed frequency of all seed tokens, as shown in Eq. \ref{eq:tokenFreq} where $\text{count}(w, S_t)$ returns the frequency of token $w$ in the stream $S$ during time slice $t$. 
These seed tokens are chosen such that they likely exhibit bursts in usage during the key moments of our sporting event data, such as ``goal'' for goals in soccer/football or hockey or ``run'' for runs scored in baseball.
This TokenBurst implementation also includes some rudimentary normalization to collapse modified words to their originals (e.g., ``gooaallll'' to ``goal'').
Many existing stream-based event detection systems use just such an approach to track specific types of events.
\begin{equation}
\label{eq:tokenFreq}
\text{freq}(t, S) = \sum_{w \in W}\text{count}(w, S_t)
\end{equation}

Since our analysis covers three separate types of sporting competitions, seed keywords should include tokens from vocabularies of each. 
We avoid separate keyword lists for each sport to provide an even comparison to the general nature of our language-agnostic technique.
The tokens for which we searched are shown in Table \ref{tab:targetTokens}.
We also used regular expressions to collapse deliberately misspelled tokens to their normal counterparts.
\begin{table}[htdp]
\footnotesize
\caption{Predefined Seed Tokens}
\begin{center}
\begin{tabular}{|p{0.75in}|p{2.0in}|}
\hline
\multicolumn{1}{|c|}{\textbf{Sport}} & \multicolumn{1}{|c|}{\textbf{Tokens}} \\ \hline
World Series & run, home, homerun \\ \hline
Super Bowl & score, touchdown, td, fieldgoal, points \\ \hline
World Cup & goal, gol, golazo, score, foul, penalty, card, red, yellow, points \\ \hline
\end{tabular}
\end{center}
\label{tab:targetTokens}
\end{table}

\subsubsection{Performance Evaluation}

Having defined LABurst, RawBurst, and TokenBurst, we evaluate these algorithms by constructing a series of receiver operating characteristic (ROC) curves across test sets of our sports data.
We then evaluate relative performance between the approaches by comparing their respective areas under the curves (AUCs) by varying the threshold parameters for each method.
In RawBurst and TokenBurst, this threshold parameter refers to $\rho$ in $\Delta_{t, k} \ge \rho$.
For our LABurst method, the ROC curve is generated by varying the minimum $\rho$ in $\widehat{\mathbbm{1}_{E'}}(S_t, t) = |\mathbf{B}_t| \ge \rho$.
The AUC of the ROC curve is useful because it is robust against imbalanced classes, which we expect to see in such an event detection task.
Then, by comparing these AUC values, we can provide an answer to \textbf{RQ1}.

\subsection{Evaluating Domain Independence}

Beyond LABurst's ability to discover and describe interesting moments, we also claim it to be domain independent.
To justify this claim, we must answer our second research question \textbf{RQ2}: can LABurst transfer models learned in one context to another one separate from its training domain and remain competitive?

Detecting key moments within sporting competitions as described above is a useful task for areas like advertising or automated highlight generation, but a more compelling and worthwhile task would be to detect higher-impact events like natural disasters.
The typical seed-token-based approach is difficult here as it is impossible to know what events are about to happen where, and a list of target keywords to detect all such events would be long and lead to false positives.
LABurst could be highly beneficial here as one need not know details like event location, language, or type.
This context presents an opportunity to evaluate LABurst in a new domain and compare it to existing work by Sakaki, Okazaki, and Matsuo \cite{Sakaki:2010:EST:1772690.1772777}.
Thus, to answer \textbf{RQ2}, we can take the LABurst model as trained on sporting events presented for \textbf{RQ1} and apply them directly to this context.

For this earthquake detection task, we compare LABurst with the TokenBurst baseline using the keyword ``earthquake,'' as in Sakaki, Okazaki, and Matsuo.
Also following Sakaki et al., we target earthquakes in Japan over the past two years and select two of the most severe: the 7.1-magnitude quake off the coast of Honshu, Japan on 25 October 2013, and a 6.5-magnitude quake off the coast of Iwaki, Japan on 11 July 2014.
Rather than generating ROC curves for this comparison, we take a more straightforward approach and compare lag between the actual earthquake event and the point in time in which the two methods detect the earthquake.
If the lag between TokenBurst and LABurst is sufficiently small, we will have good evidence for an affirmative answer to \textbf{RQ2}.

\section{Data Collection}

While the algorithms described herein are general and can be applied to any sufficiently active social media stream, the ease with which one can access and collect Twitter data makes it an attractive target for our research.
To this end, we leveraged two existing Twitter corpora and created our own corpus of tweets from Twitter's 1\% public sample stream\footnote{https://dev.twitter.com/streaming/reference/get/statuses/sample}.
This new corpus was created using the twitter-tools library\footnote{https://github.com/lintool/twitter-tools} developed for evaluations at the NIST Text Retrieval Conferences (TRECs).
In collecting from Twitter's public sample stream, we connect to the Twitter API endpoint (provide \textbf{no filters}), and retrieve a sampling of 1\% of all public tweets, which yields approximately 4,000 tweets per minute.

The two existing corpora we used were the Edinburgh Corpus \cite{Petrovic:2010:ETC:1860667.1860680}, which covered the 2010 NFL division championship game, and an existing set of tweets pulled from Twitter's firehose source targeted at Argentina during November of 2012, which covered the four Premier League soccer games.
All remaining data sets were extracted from Twitter's sample stream over the course of October 2013 to July 2014.

Where possible, for each event (both sporting and earthquake), we recorded all tweets from the 1\% stream starting an hour before the target event and ending an hour after the event, yielding over 15 million tweets.
Table \ref{tab:tweetCounts} shows the breakdown of tweets collected per event.
From these tweets,  we extracted $1,109$ positive (i.e., known bursty) samples and $43,037$ negative samples for a total of $44,146$ data points.

\begin{table}[htdp]
\footnotesize
\caption{Per-Event Tweet Counts}
\begin{center}
\begin{tabular}{|p{2in}|c|}
\hline
\multicolumn{1}{|c|}{\textbf{Event}} & \textbf{Tweet Count} \\ \hline
\multicolumn{2}{|c|}{\textbf{Training Data}}  \\ \hline
2010 NFL Division Championship & 109,809 \\ \hline
2012 Premier League Soccer Games & 1,064,040 \\ \hline
2014 NHL Stanley Cup Playoffs & 2,421,065 \\ \hline
2014 NBA Playoffs & 500,170 \\ \hline
2014 Kentucky Derby Horse Race & 233,172 \\ \hline
2014 Belmont Stakes Horse Race & 226,160 \\ \hline
2014 FIFA World Cup Stages A+B & 5,867,783 \\ \hline

\multicolumn{2}{|c|}{\textbf{Testing Data}}  \\ \hline
2013 MLB World Series Game 5  & 1,052,852 \\ \hline
2013 MLB World Series Game 6  & 1,026,848 \\ \hline
2013 Honshu Earthquake  & 444,018 \\ \hline
2014 NFL Super Bowl & 1,024,367 \\ \hline
2014 FIFA World Cup Third Place & 809,426 \\ \hline
2014 FIFA World Cup Final & 1,166,767 \\ \hline
2014 Iwaki Earthquake  & 358,966 \\ \hline

\multicolumn{1}{|r|}{\textbf{Total}} &  16,305,443  \\ \hline
\end{tabular}
\end{center}
\label{tab:tweetCounts}
\end{table}

\section{Experimental Results}
\label{sect:results}

\subsection{Setting Model Parameters}

Prior to carrying out the experiments described above, we first needed appropriate parameters for window sizes and LABurst's classifiers.
For LABurst's slice size $\delta$, window size $\omega$, and $k$ previous window parameters, preliminary experimentation yielded acceptable results with the following: $\delta = 60$ seconds, $\omega = 180$ seconds, and $k=10$.
We used these $\delta$ and $k$ parameters in both RawBurst and TokenBurst as well.

Regarding LABurst's classifier implementations, we used the Scikit-learn\footnote{http://scikit-learn.org/} Python package for SVMs and RFs as well as an implementation of the ensemble classifier AdaBoost, each of which provided a number of hyperparameters to set.
For SVMs, the primary hyperparameter is the type of kernel to use, and initial experiments showed SVMs with linear kernels performed poorly.
We then applied principal component analysis to reduce the training data's dimensionality to a three-dimensional space for visualization.
The resulting visualization showed a decision boundary more consistent with a sphere rather than a clear linear plane, motivating our choice of the radial basis kernel (RBF). 

For the remaining hyperparameters, we constructed separate parameter grids for SVMs and RFs and performed a distributed grid search.
The grid for SVM's two parameters, cost $c$ and kernel coefficient $\gamma$, covered powers of two such that ${c, \gamma} = 2^x$, $x \in [-2, 10]$.
RF parameters were similar for the number of estimators $n$ and feature count $c'$ such that $n = 2^x$, $x \in [0, 10]$ and $c' = 2^y$, $y \in [1, 12]$.

Each parameter set was scored using the AUC metric across a randomly split 10-fold cross-validation set, with the best scores determining the parameters used in our ensemble.
We then combined the two classifiers using Scikit-learn's AdaBoost implementation, yielding the results shown in Table \ref{tab:scores}.
These grid search results show RFs perform better than SVMs, and the AdaBoost ensemble outperforms each individual classifier.

\begin{table}[htdp]
\footnotesize
\caption{Per-Classifier Hyperparameter Scores}
\begin{center}
\begin{tabular}{|c|c|c|c|c|}
\hline
\textbf{Classifier} & \textbf{Params} & \textbf{ROC-AUC} \\ \hline
SVM & kernel $=$ RBF, & 87.48\% \\ 
& $c=64,$ &  \\ 
& $\gamma=0.0625$ & \\ \hline
RF & trees = 1024, & 88.35\% \\
& features = 2 &  \\ \hline
AdaBoost & estimators = 2 & 89.84\% \\ \hline
\end{tabular}
\end{center}
\label{tab:scores}
\end{table}

\subsection{Ablation Study}

Given the various features from both our own development and related works, we should address the relative values or importance of each feature to our task.
To answer this question, we performed an ablation study with a series of classifiers, each excluding a single feature set.
Each degenerate classifier was then compared with the full AdaBoost classifier using the same 10-fold cross-validation strategy as above.
Table \ref{tab:ablation} shows each model's AUC and its difference with that of the full model.
These results suggest the regression and entropy features contribute the most while the average difference features seem to hinder performance.

\begin{table}[htdp]
\footnotesize
\caption{Ablation Study Results}
\begin{center}
\begin{tabular}{|c|c|c|}
\hline
\textbf{Feature Sets} & \textbf{ROC-AUC} & \textbf{Difference} \\ \hline
AdaBoost, All Features	&	89.84\%	&	--	\\ \hline
Without Regression	&	87.79\%	&	-2.05	\\ \hline
Without Entropy	&	87.94\%	&	-1.90	\\ \hline
Without TF-IDF		&	88.85\%	&	-0.99	\\ \hline
Without TF-PDF	&	89.00\%	&	-0.84	\\ \hline
Without Density		&	89.07\%	&	-0.77	\\ \hline
Without InterArrival	&	89.46\%	&	-0.38	\\ \hline
Without BursT		&	89.52\%	&	-0.31	\\ \hline
Without Average Difference	&	90.56\%	&	0.72	\\ \hline
\end{tabular}
\end{center}
\label{tab:ablation}
\end{table}

\begin{figure*}[tb]
\centering
\begin{subfigure}[b]{0.32\textwidth}
\centering
\includegraphics[width=2.2in]{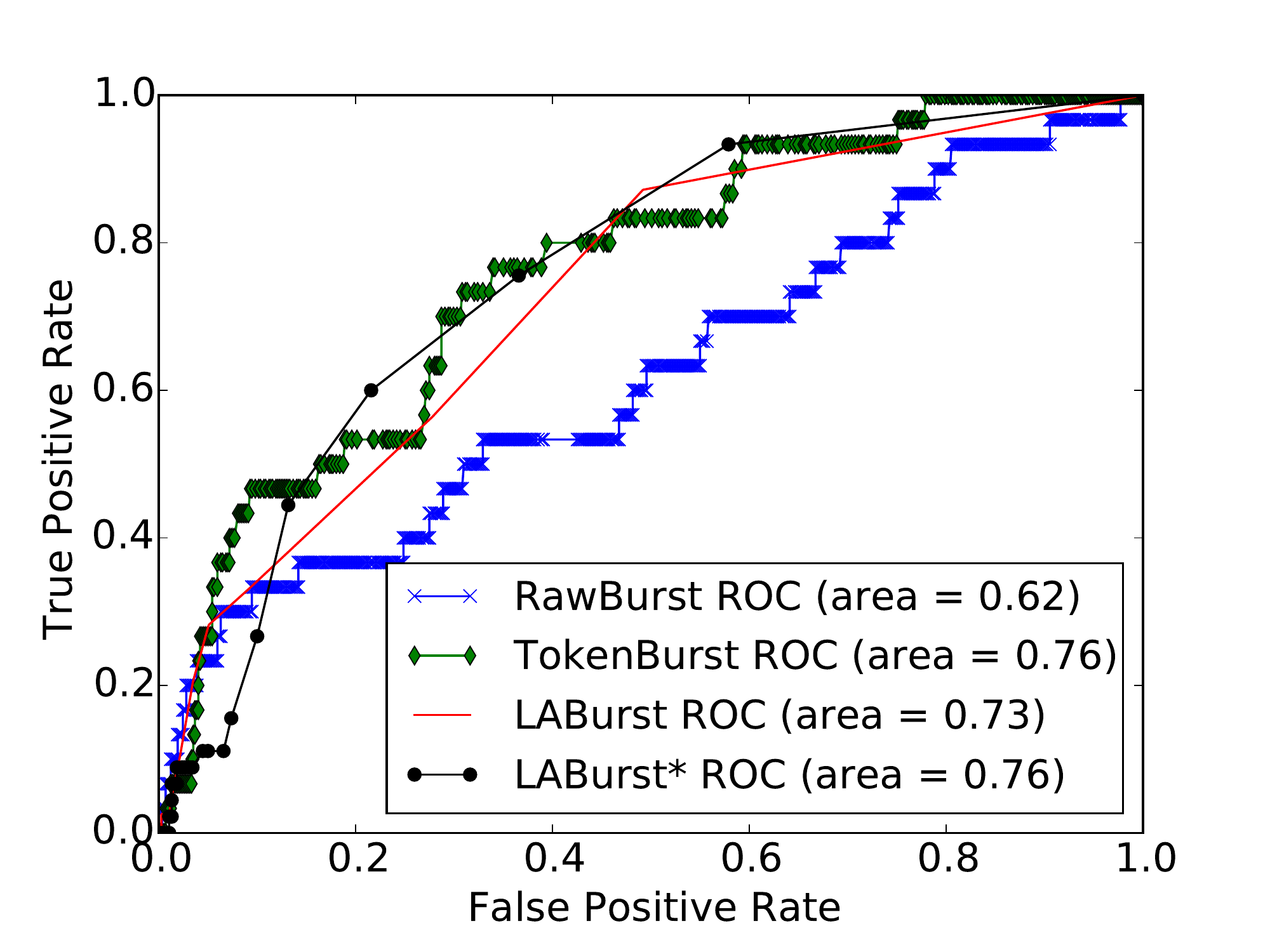}
\caption{2013 World Series}
\label{fig:roc2013WorldSeries}
\end{subfigure}
\begin{subfigure}[b]{0.32\textwidth}
\centering
\includegraphics[width=2.2in]{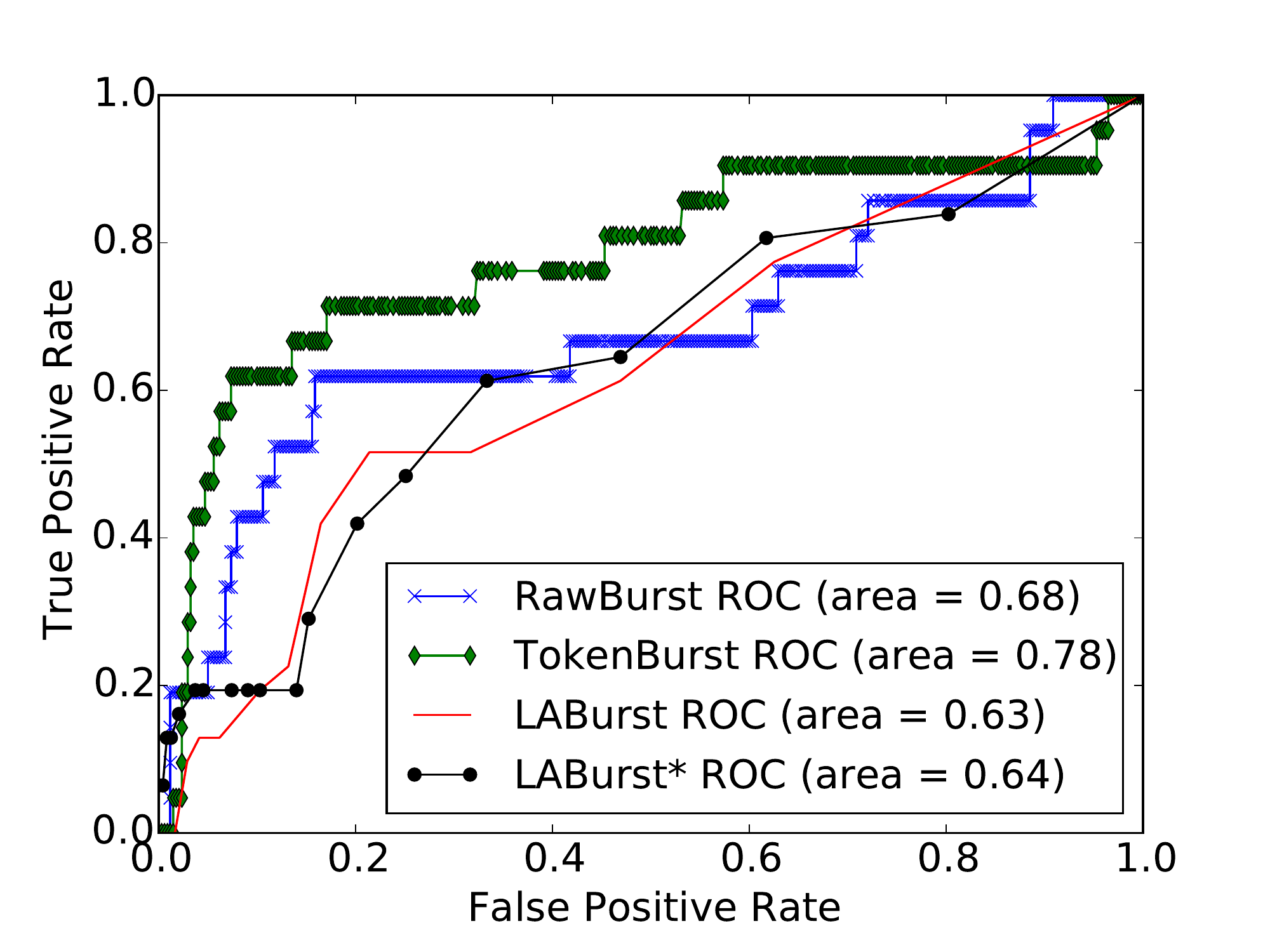}
\caption{2014 Super Bowl}
\label{fig:roc2014SuperBowl}
\end{subfigure}
\begin{subfigure}[b]{0.32\textwidth}
\centering
\includegraphics[width=2.2in]{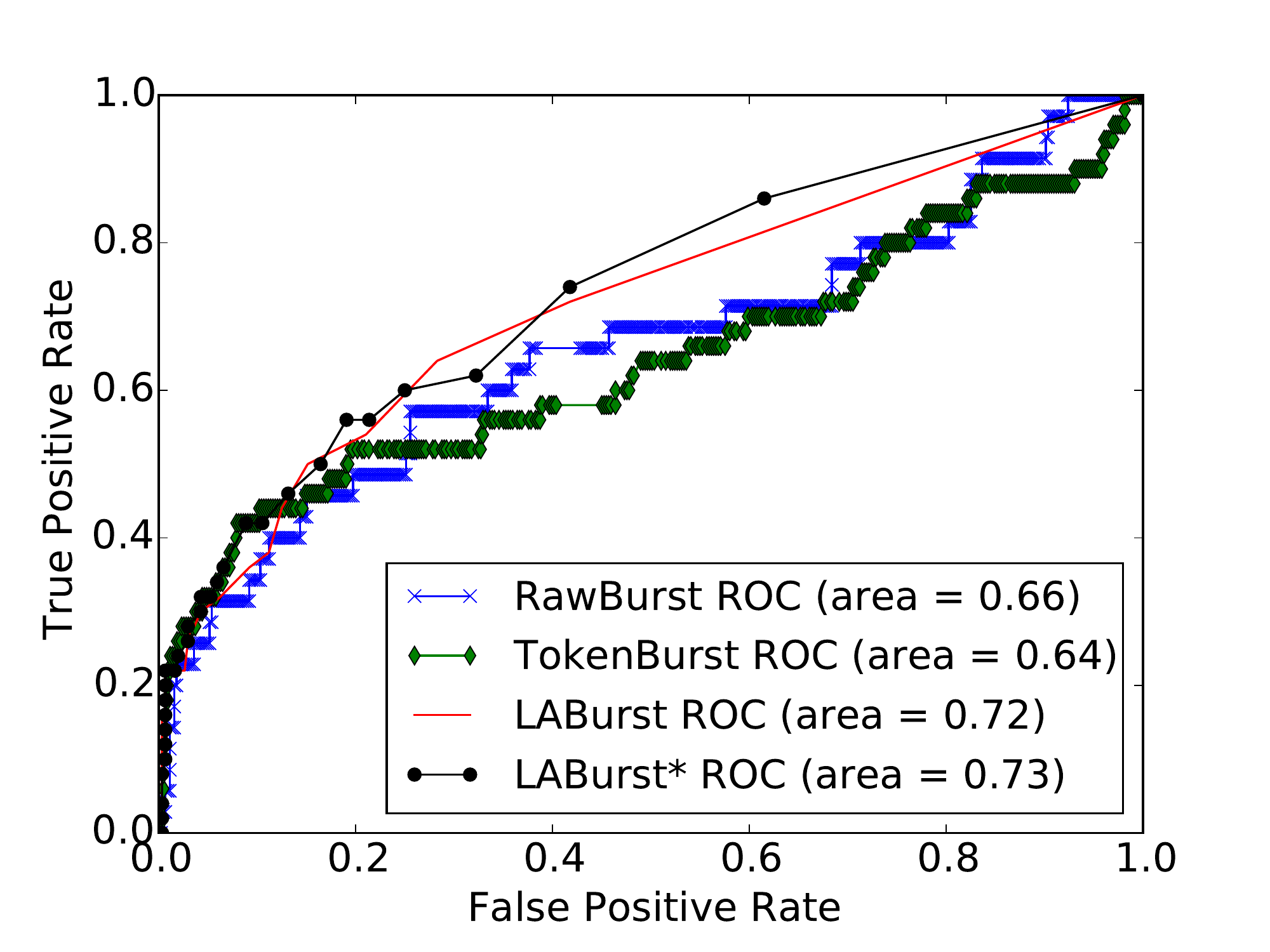}
\caption{2014 World Cup}
\label{fig:roc2014WorldCup}
\end{subfigure}
\caption{Per-Sport ROC Curves}
\label{fig:joinedPerf}
\end{figure*}

\subsection{Event Discovery Results}

To restate, the first research question (\textbf{RQ1}) posed in this work is whether LABurst can perform as well as existing methods in detecting key moments.
For convenience, we focus on sporting competitions, specifically training across several sporting events as outlined in Tables \ref{tab:eventStats} and \ref{tab:tweetCounts}, and testing on the final two games of the 2013 MLB World Series, the 2014 NFL Super Bowl, and the final two matches of the 2014 FIFA World Cup.
Prior to presenting comprehensive results, we first examine performance curves for each sporting competition, as shown in Figure \ref{fig:joinedPerf}.
Each graph in Figure \ref{fig:joinedPerf} corresponds to a particular sport, with the blue and green lines showing the ROC curves for RawBurst and TokenBurst respectively.
The red line shows the ROC curve for the LABurst model trained using all features, whereas the black line illustrates the LABurst model trained using all but the average difference feature set.
We refer to this restricted version as LABurst*.

For the 2013 World Series, RawBurst's AUC is 0.62, TokenBurst's is 0.76, LABurst achieves 0.73, and LABurst* yields 0.76.
From \ref{fig:roc2013WorldSeries}, the two LABurst models clearly dominate RawBurst and exhibit performance on par with TokenBurst.
During the Super Bowl, RawBurst and TokenBurst achieve an AUC of 0.68 and 0.78 respectively, while LABurst and LABurst* perform worse with an AUC of 0.63 and 0.64, as shown in Figure \ref{fig:roc2014SuperBowl}.
During the 2014 World Cup, both LABurst and LABurst* (AUC = 0.72 and 0.73) outperformed both RawBurst (AUC = 0.66) and TokenBurst (AUC = 0.64), as seen in Figure \ref{fig:roc2014WorldCup}.

\subsection{Composite Results}

\begin{figure}[htbp]
\begin{center}
\includegraphics[width=2.3in]{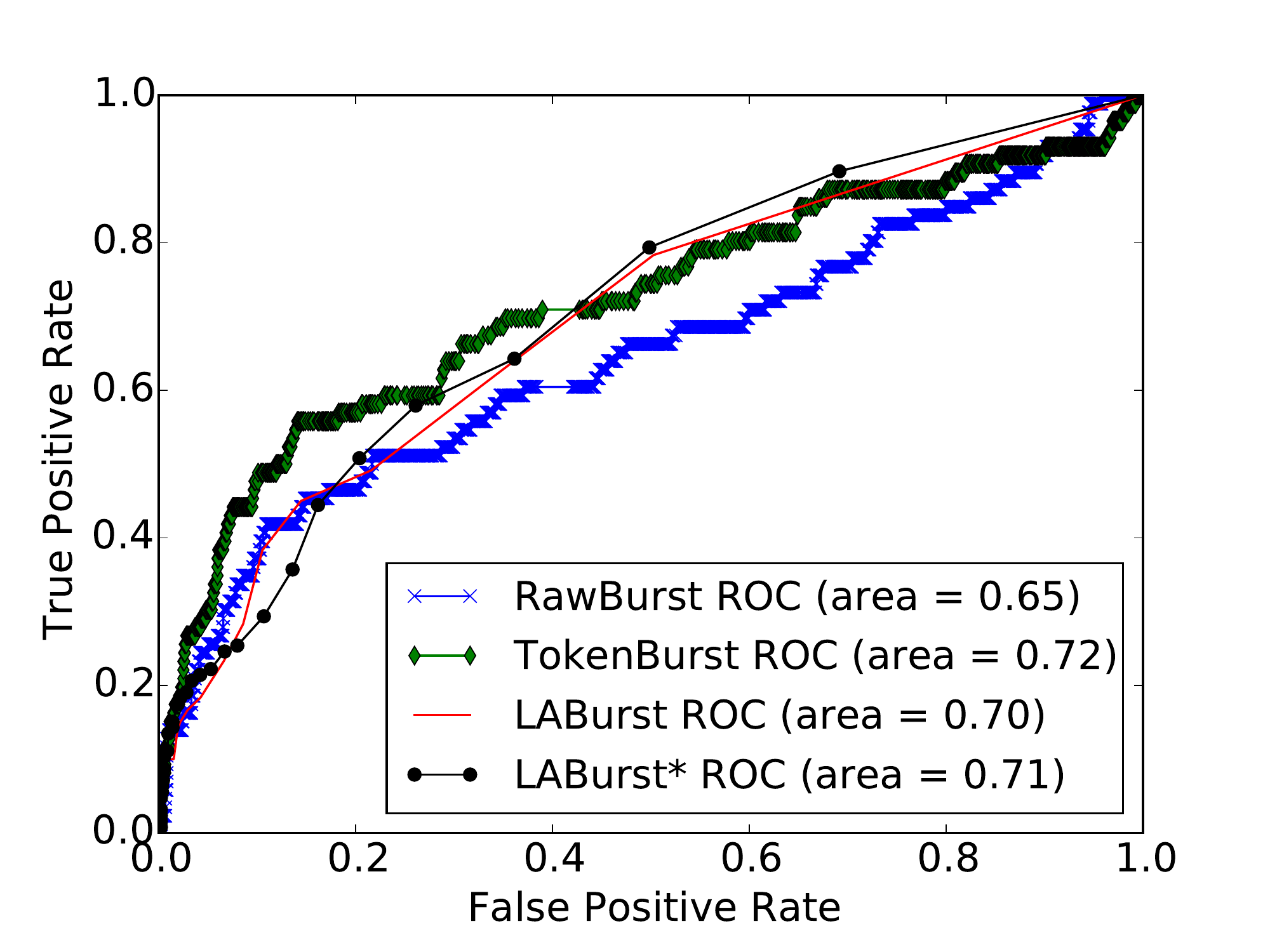}
\caption{Composite ROC Curves}
\label{fig:rocComprehensive}
\end{center}
\end{figure}

To compare comprehensive performance, we look to Figure \ref{fig:rocComprehensive}, which shows ROC curves for all three methods across all three testing events.
From this figure, we see LABurst (AUC=0.7) and LABurst* (AUC=0.71) both outperform RawBurst (AUC=0.65) and perform nearly as well as TokenBurst (AUC=0.72).
Given these results, one can answer \textbf{RQ1} in that, yes, LABurst is competitive with existing methods.

More interestingly, assuming equal cost for false positives and negatives and optimizing for the largest difference between true positive rate (TPR) and false positive rate (FPR), TokenBurst shows a TPR of 0.56 and FPR of 0.14 with a difference of 0.42 at a threshold value of 13.2.
LABurst, on the other hand, has a TPR of 0.64 and FPR of 0.28 with a difference of 0.36 at a threshold value of 2.
From these values, we see LABurst achieves a higher true positive rate at the cost of a higher false positive rate. 
This effect is possibly explained by the domain-specific nature of our test set and TokenBurst implementation, as discussed in more detail in Section \ref{sect:SuperBowlDeficiency}.

\begin{figure*}[hbtp]
\centering
\begin{subfigure}[b]{0.4\textwidth}
\includegraphics[width=2.25in]{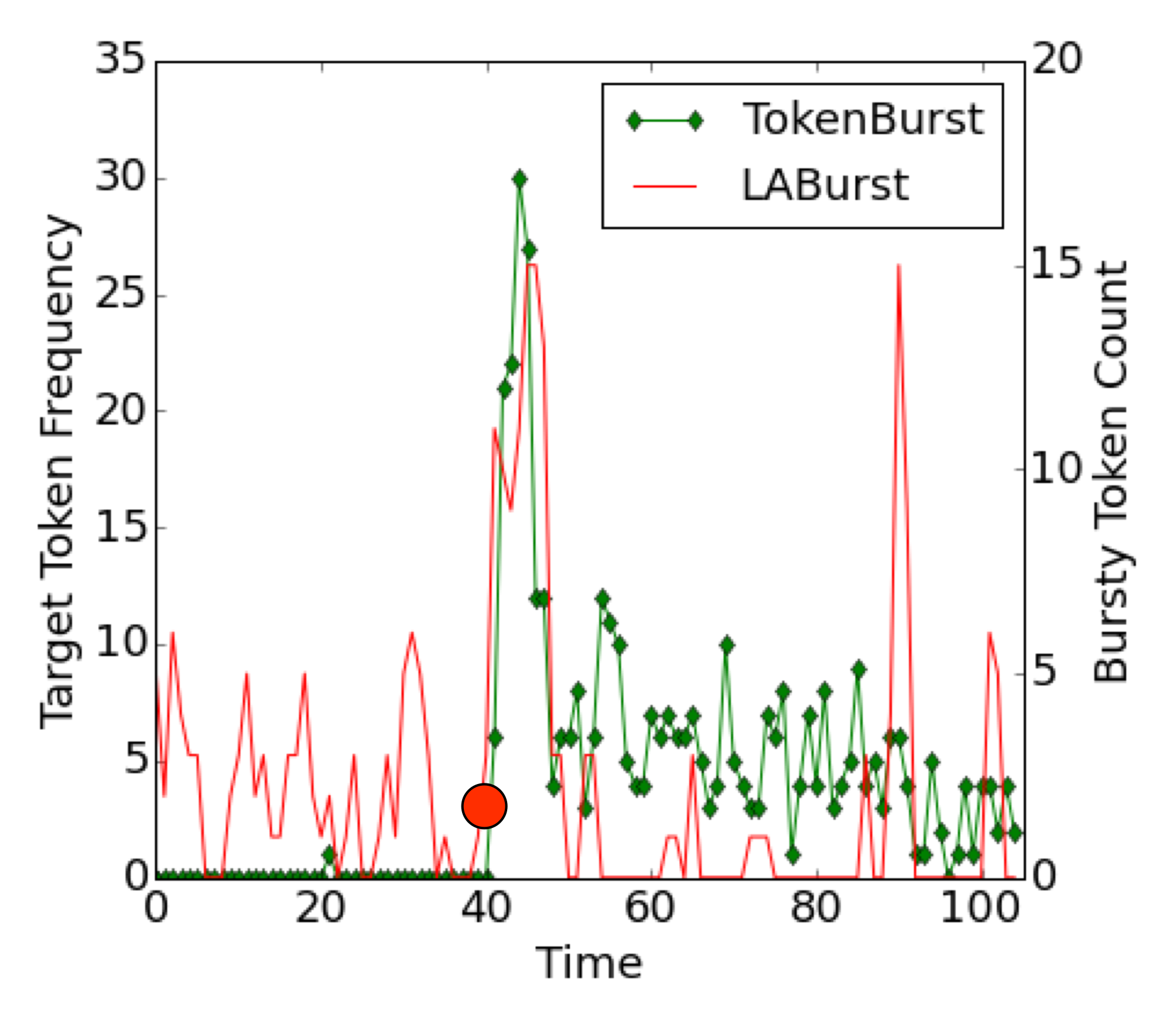}
\caption{Honshu, Japan Earthquake - 25 October 2013}
\label{fig:2013Japan}
\end{subfigure}
\begin{subfigure}[b]{0.4\textwidth}
\includegraphics[width=2.25in]{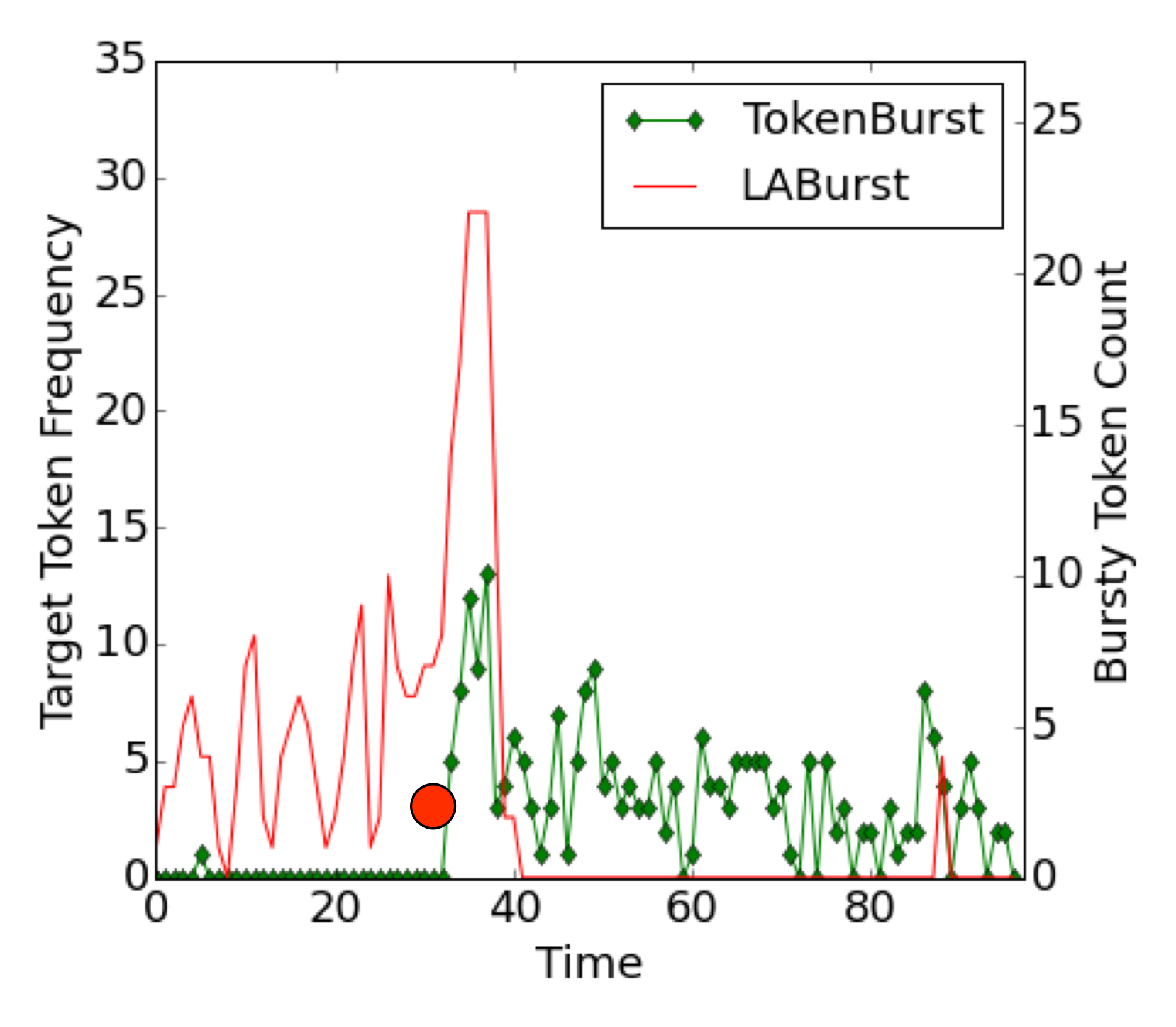}
\caption{Iwaki, Japan Earthquake - 11 July 2014}
\label{fig:2014Japan}
\end{subfigure}
\caption{Japanese Earthquake Detection}
\label{fig:joinedEarthquake}
\end{figure*}

\subsection{Earthquake Detection}

Our final research question (\textbf{RQ2}) seeks to determine if adapting LABurst's models, as trained on using sporting events listed in Tables \ref{tab:eventStats} and \ref{tab:tweetCounts}, can compete with existing techniques in a different domain.
We explored this adaptation by applying the sports-trained LABurst classifier to Twitter data surrounding known earthquake events in Japan in 2013 and 2014.

Figures \ref{fig:2013Japan} and \ref{fig:2014Japan} show the detection curves for both methods for the 2013 and 2014 earthquakes respectively; the red dots indicate the earthquake times as reported by the United States Geological Survey (USGS).
The left vertical axis for each figure reports the frequency of the ``earthquake'' token, and the right axis shows the number of tokens classified as bursty by LABurst.
From the TokenBurst curve, one can see the token ``earthquake'' sees a significant increase in usage when the earthquake occurs, and LABurst experiences a similar increase simultaneously.
It is worth noting that LABurst exhibits bursts prior to the earthquake event, but these peaks \emph{are unrelated} to the earthquake event since LABurst does not differentiate between the earthquake and other high-impact events that could be happening on Twitter. 
In addition, the peak occurring about 50 minutes after the earthquake on 25 October 2013 potentially represents an aftershock event\footnote{http://ds.iris.edu/spud/aftershock/9761021}.
Given the minimal lag between LABurst and TokenBurst's detection, we have shown LABurst is effective in cross-domain event discovery (\textbf{RQ2}).

One can now ask what tokens we identified as bursting when the earthquakes occurred.
Many of the tokens are in Japanese, and tokens at the peak of the earthquake events are shown in Table \ref{tab:japanTokens}.
We also extracted several tweets that contain the highest number of these tokens for the given time period, a selection of which include, \begin{CJK}{UTF8}{\myfont}``地震だあああああああああああああああああああああ,'' ``今回はチート使ってないから地震わからなかった,'' and ``地震だー.'' \end{CJK}
Google Translate\footnote{http://translate.google.com} translates these tweets as ``Ah ah ah ah ah ah ah ah ah Aa's earthquake,'' ``I did not know earthquake because not using cheat this time,'' and ``Over's earthquake'' respectively.
\begin{table}[htdp]
\footnotesize
\caption{Discovered Bursty Tokens}
\begin{center}
\begin{tabular}{|p{1.45in} | p{1.45in} |}
\hline
\multicolumn{1}{|c|}{\textbf{Earthquake}} & \multicolumn{1}{|c|}{\textbf{Bursty Tokens}} \\ \hline
Honshu, Japan -- 25 October 2013 & \c{c}\begin{CJK}{UTF8}{\myfont} 丈, 地, 夫, 怖, 波, 注, 津, 源, 福, 震 \end{CJK} \\ \hline
Iwaki, Japan -- 11 July 2014 & \begin{CJK}{UTF8}{\myfont} び, ゆ, ビビ, 地, 怖, 急, 福, 警, 速, 震 \end{CJK} \\ \hline
\end{tabular}
\end{center}
\label{tab:japanTokens}
\end{table}

\section{Analysis}
\label{sect:analysis}

In comparing LABurst with the baseline techniques, it is important to note the strengths and weaknesses of each baseline: RawBurst requires no prior information but provides little in the way of semantic information regarding detected events, while TokenBurst provides this semantic information at the cost of missing unknown tokens or significant events that do not conform to its prior knowledge.
LABurst attempts to combine these two approaches by supporting undirected event discovery while yielding insight into these moments by tagging relevant bursting tokens.

\subsection{Identifying Event-Related Tokens}

As mentioned, where the baselines sacrifice either insight or flexibility, LABurst jointly attacks these problems and yields event-related tokens automatically.
These tokens may include misspellings, colloquialisms, and language-crossing tokens, which makes them hard to know a priori.
The 2014 World Cup provides an illustrative case for such unexpected tokens given its enormous viewership: many Twitter users of many different languages are likely tweeting about the same event.
Table \ref{tab:burstyTokens} shows a selection of events from the final two World Cup matches and a subset of those tokens classified as bursting during the events (one should note the list is not exhaustive owing to formatting and space constraints).

\begin{table}[htdp]
\footnotesize
\caption{Tokens Classified as Busting During Events}
\begin{center}
\begin{tabular}{|p{0.75in}|p{0.7in}| p{1.45in} |}
\hline
\multicolumn{1}{|c|}{\textbf{Match}} & \multicolumn{1}{|c|}{\textbf{Event}} & \multicolumn{1}{|c|}{\textbf{Bursty Tokens}} \\ \hline
Brazil v. Netherlands, 12 July 2014 & Netherlands' Van Persie scores a goal on a penalty at 3', 1-0 & 0-1, 1-0, 1:0, 1x0, card, goaaaaaaal, goal, gol, goool, holandaaaa, k\i{}rm\i{}z\i{}, pen, penal, penalti, p\^{e}nalti, persie, red \\ \hline
Brazil v. Netherlands, 12 July 2014 & Brazil's Oscar get's a yellow card at 68' & dive, juiz, penalty, ref \\ \hline
Germany v. Argentina, 13 July 2014 & Germany's G\"{o}tze scores a goal at 113', 1-0 & goaaaaallllllll, goalllll, godammit, goetze, gollllll, gooooool, gotze, gotzeeee, g\"{o}tze, nooo, yessss, \begin{CJK}{UTF8}{\myfont} ドイツ\end{CJK} \\ \hline
\end{tabular}
\end{center}
\label{tab:burstyTokens}
\end{table}

Several interesting artifacts emerge from this table, first of which is that one can get an immediate sense of what happened in the detected moment from tokens our algorithm presents. 
For instance, the prevalence of the token ``goal'' and its variations clearly indicate a team scored in the first and third events in Table \ref{tab:burstyTokens}; similarly, bursting tokens associated with the middle event regarding Oscar's yellow card reflect his penalty for diving.
Beyond the pseudo event description put forth by the identified tokens, references to diving and specific player/team names in the first and third events are also of significant interest.
In the first event, one can infer that the Netherlands scored since ``holandaaaa'' is flagged along with ``persie'' for the Netherlands' player, Van Persie, and likewise for Germany's G\"{o}tze in the third event (and the accompanying variations of his name).
These tokens would be difficult to capture beforehand as TokenBurst would require, and such tokens would likely not be related to every event or every type of sporting event.

Finally, the last artifact of note is that the set of bursty tokens displayed includes tokens from several different languages: English for ``goal'' and ``penalty,'' Spanish for ``gol'' and ``penal,'' Brazilian Portuguese for ``juiz'' (meaning ``referee''), as well as the Arabic for ``goal'' and Japanese for ``Germany.''
Since these words are semantically similar but syntactically distinct, typical normalization schemes could not capture these connections.
Instead, capturing these words in the baseline would require a pre-specified keyword list in all possible languages or a machine translation system capable of normalizing within different languages (to collapse ``goool'' down to ``gol'' for example).

\subsection{Discovering Unanticipated Moments}
\label{sect:eventDiscovery}

\begin{figure}[hbtp]
\begin{center}
\includegraphics[width=2.75in]{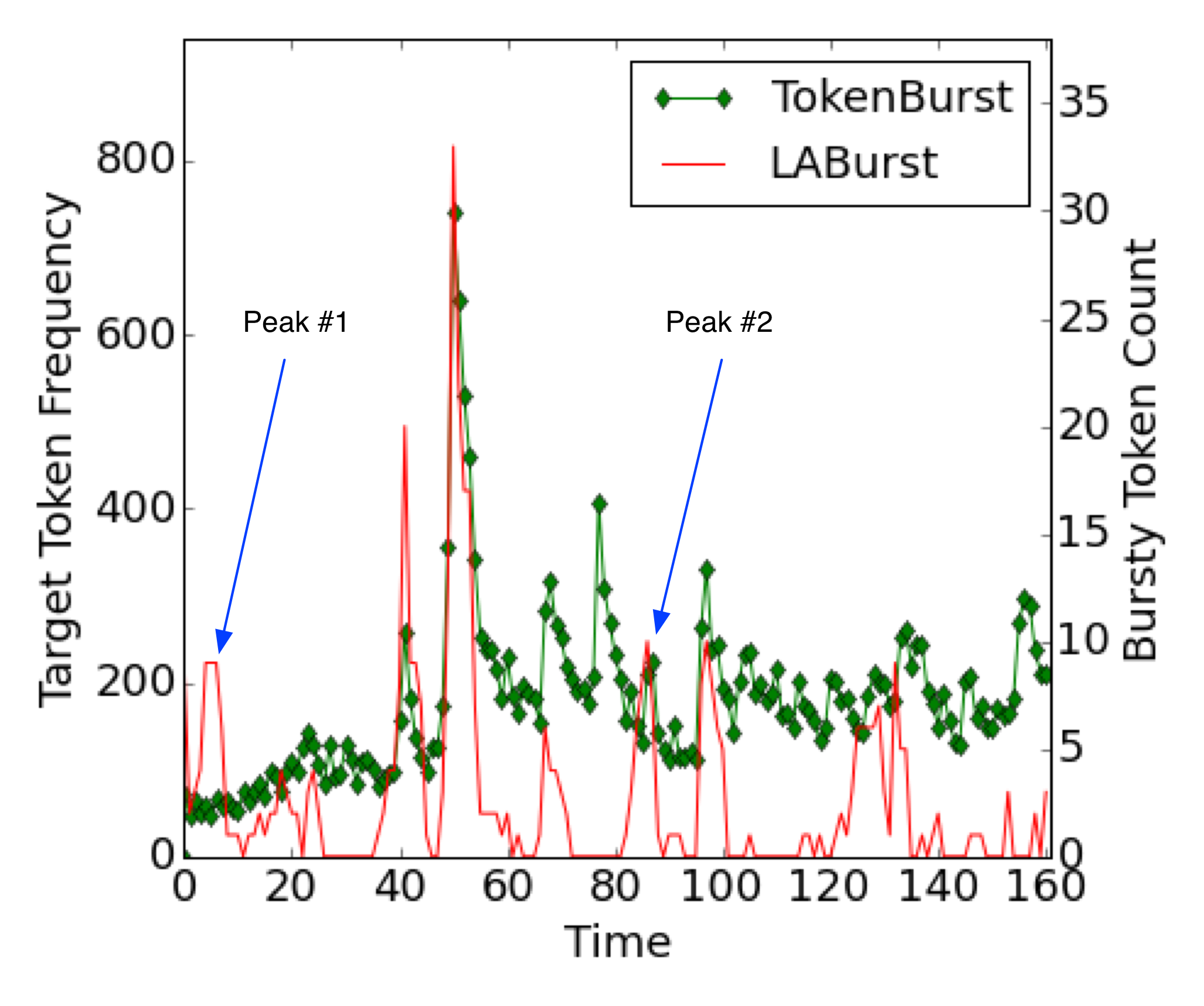}
\caption{Baseline and LA Bursty Frequencies}
\label{fig:worldCupFreqs}
\end{center}
\end{figure}

Results show LABurst is competitive with the domain-specific TokenBurst, but TokenBurst's specificity makes it unable to detect unanticipated moments, and we can see instances of such omissions in the last game of World Cup.
Figure \ref{fig:worldCupFreqs} shows target token frequencies for TokenBurst in green and LABurst's volume of bursty tokens in red.
From this graph, we can see the first instance in Peak \#1 where LABurst exhibits a peak  missed by TokenBurst.
Tokens appearing in this peak include ``puyol,'' ``gisele,'' and ``bundchen,'' which correspond to former Spanish player Carles Puyol and model Gisele Bundchen, who presented the World Cup trophy prior to the match.
While not necessarily a sports-related event, many viewers were interested in the trophy reveal, making it a key moment.
At peak \#2, slightly more than eighty minutes into the data (which is sixty minutes into the match), LABurst sees another peak otherwise inconspicuous in the TokenBurst curve.
Upon further exploration, tokens present in this peak refer to Argentina's substituting Ag\"{u}ero for Lavezzi at the beginning of the match's second half.

\subsection{Addressing the Super Bowl}
\label{sect:SuperBowlDeficiency}

While LABurst performs as well as the domain-specific TokenBurst algorithm in both the World Series and World Cup events, one cannot ignore its poor performance during the Super Bowl.
Since LABurst is both language agnostic and domain independent, it likely detects additional high-impact events outside of the game start/end, score, and penalty events present in our ground truth.
For instance, during the Super Bowl, spectators tweet about moments beyond sports plays: they tweet about the half-time show, commercials, and massive power outages.
Since our ground truth disregards such moments, LABurst's higher false-positive rate is less surprising, and TokenBurst's superior performance might result from its specificity in domain knowledge with respect to the ground truth (i.e., both include only sports data).
Hence, LABurst's ability to detect unanticipated moments potentially penalizes it in domain-specific tasks.

LABurst's propensity towards more organic moments of interest becomes obvious when we inspect the tokens LABurst identified when it detected a large burst early on that TokenBurst missed.
Approximately four minutes before the game started (and therefore before when TokenBurst would detect any event), LABurst saw a large burst with tokens like ``joe'', ``namath'', ``fur'', ``coat'', ``pimp'', ``jacket'', ``coin'', and ``toss''.
As it turns out, Joe Namath, an ex football star, garnered significant attention from fans when he tossed the coin to decide which team would get first possession. 
Since neither our ground truth data nor TokenBurst's domain knowledge captured this moment, LABurst's detection is counted as a false positive much like the trophy presentation during the World Cup. 

\section{Limitations and Extensions}

The approach adopted herein is fundamentally limited regarding tracking potentially interesting events that do not garner mass awareness on social media. 
Since the LABurst presupposes significant bursts in activity during key moments, if only a few people are participating in or following an event, LABurst will likely be unable to detect moments in that event.
This effect is clear in applying LABurst to regular season baseball games: since Major League Baseball sees over 2,400 games in a season, experiments showed too few viewers were posting messages to Twitter during these games to generate any significant burst.
As a result, many key moments in these games are exceedingly difficult to capture via burst detection.

This deficiency leads to a potential opportunity, however, in combining domain knowledge with LABurst's domain-agnostic foundations.
For example, one could apply domain-specific filters to the Twitter stream prior to LABurst in the detection pipeline.
Since LABurst uses relative frequencies to identify bursts, this pre-filtering step should amplify the signal of potentially bursty tokens in the stream and increase LABurst's likelihood of detecting them.
Returning to the baseball example, one could use domain information to filter the Twitter stream to contain only relevant tweets, and the baseball-specific key moments should become more apparent.

In a more interesting case, this domain knowledge could be applied as events are discovered and allow LABurst to provide more insight into those events as they unfold.
Examples where such an approach could be used include hurricanes, where one can know the name of the hurricane and its approximate area of landfall, filter the Twitter stream accordingly, and then use LABurst to track the unanticipated moments that occur once the storm hits.
One could apply a similar approach in the early hours of political protests or mass unrest to track events that may not be covered by mainstream news outlets (e.g., in oppressive regimes where media is controlled).
Additional knowledge such as geolocation data could also be integrated into these stream filters to increase LABurst's moment discovery capabilities further.


\section{Conclusions}
\label{sect:conlusions}

Revisiting motivations, this research sought to demonstrate whether LABurst, a \textbf{streaming}, \textbf{language-agnostic}, \textbf{burst-centric} algorithm, can discover key moments from unfiltered social media streams (specifically Twitter's public sample stream).
Our results show temporal features can identify bursty tokens and, using the volume of these tokens as an indicator, we can discover key moments across a collection of disparate sporting competitions.
This approach's performance is \textbf{competitive with existing baselines}.
Furthermore, these sports-trained models are \textbf{adaptable to other domains} with a level of performance exceeding a simple time series baseline and rivaling a domain-specific method.
LABurst's performance relative to the domain-specific baseline shows this method's potential given its \textbf{omission of manual keyword selection} and prior knowledge.

Beyond this comparison, our approach also offers notable flexibility in identifying bursting tokens  across language $\;$ boundaries and in supporting event description; that is, we can get a sense of the occurring event by inspecting bursty tokens returned by LABurst.
These features combine to form a capable tool for discovering unanticipated moments of high interest, regardless of language.
This technique is particularly useful for journalists and first responders, who have a vested interest in rapidly identifying and understanding high-impact moments, even if a journalist or aid worker is not physically present to observe the event.
Possibilities also exist to combine LABurst with other domain-specific solutions and yield insight into unanticipated events, events missed by existing approaches, or events that might otherwise be lost in the noise.


\section{Acknowledgments}
This work was supported in part by the National Science Foundation under CNS-1405688 \cite{Lin:2014ty}. 
Any opinions, findings, conclusions, or recommendations expressed are those of the authors and do not necessarily reflect the views of the sponsors.
This work also made use of the Open Science Data Cloud $\;$ (OSDC), which is an Open Cloud Consortium (OCC) - sponsored project. 
The OSDC is supported in part by grants from Gordon and Betty Moore Foundation and the National Science Foundation and major contributions from OCC members like the University of Chicago. 

%
\bibliographystyle{abbrv}
\bibliography{arXiv15}  

\end{document}